\def \AAP #1 #2 {{\em Astron. Astrophys.\/} {\bf #1}, #2}
\def \AAL #1 #2 {{\em Astron. Astrophys. Lett.\/} {\bf #1}, L#2}
\def \AAR #1 #2 {{\em Astron. Astrophys. Rev.\/} {\bf #1}, #2}
\def \AAS #1 #2 {{\em Astron. Astrophys. Suppl. Ser.\/} {\bf #1}, #2}
\def \AJ #1 #2 {{\em Astron. J.\/} {\bf #1}, #2}
\def \ANNREV #1 #2 {{\em Ann. Rev. Astron. Astrophys.\/} {\bf #1}, #2}
\def \APJ #1 #2 {{\em Astrophys. J.\/} {\bf #1}, #2}
\def \APJL #1 #2 {{\em Astrophys. J. Lett.\/} {\bf #1}, L#2}
\def \APJS #1 #2 {{\em Astrophys. J. Suppl.\/} {\bf #1}, #2}
\def \APSS #1 #2 {{\em Astrophys. Space Sci.\/} {\bf #1}, #2}
\def \ASR #1 #2 {{\em Adv. Space Res.\/} {\bf #1}, #2}
\def \BAIC #1 #2 {{\em Bull. Astron. Inst. Czechosl.\/} {\bf #1}, #2}
\def \JSQRT #1 #2 {{\em J. Quant. Spectrosc. Radiat. Transfer\/} {\bf #1}, #2}
\def \MN #1 #2 {{\em Mon. Not. R. Astr. Soc.\/} {\bf #1}, #2}
\def \MEM #1 #2 {{\em Mem. R. Astr. Soc.\/} {\bf #1}, #2}
\def \PLR #1 #2 {{\em Phys. Lett. Rev.\/} {\bf #1}, #2}
\def \PASJ #1 #2 {{\em Publ. Astron. Soc. Japan\/} {\bf #1}, #2}
\def \PASP #1 #2 {{\em Publ. Astr. Soc. Pacific\/} {\bf #1}, #2}
\def \NAT #1 #2 {{\em Nature\/} {\bf #1}, #2}
\def \SAIT #1 #2 {{\em Mem.\ Soc.\ Astron.\ It.\/} {\bf #1}, #2}
\def \MESS #1 #2 {{\em The Messenger\/} {\bf #1}, #2}
\def \ASTRNACH #1 #2 {{\em Astron. Nach.\/} {\bf #1}, #2}

\documentstyle{BSAXPwork}
\input epsf.sty
\begin{opening}
\title{Observation of Soft Gamma Repeaters with BeppoSAX}
\author{Marco Feroci$^{1}$}
\institute{$^1$Istituto di Astrofisica Spaziale e Fisica
Cosmica,\\
Consiglio Nazionale delle Ricerche, Rome, Italy}
\date{} % DO NOT INSERT ANY DATE HERE !!!
\end{opening}

\begin{document}

%\oddpagefooter{\sf Pulsars, AXPs and SGRs BeppoSAX Workshop}{}{\thepage}
%\evenpagefooter{\thepage}{}{\sf BeppoSAX Pulsar Workshop}
\oddpagefooter{}{}{} % LEAVE AS IT IS !
\evenpagefooter{}{}{} % LEAVE AS IT IS !
\medskip  % LEAVE AS IT IS !

\begin{abstract}
In this paper I will briefly review what are, in my view, the main
contributions of BeppoSAX to the understanding of the class of
sources known as Soft Gamma Repeater. These enigmatic sources were
firmly identified as steady pulsars just during the operating
lifetime of BeppoSAX. All the instruments onboard BeppoSAX have at
some level contributed in this field with specific observations,
always allowing high quality - sometimes unprecedented - studies
of the quiescent counterparts or the bursting behavior of these
sources. I will try to stress the results that were uniquely
achieved by BeppoSAX and identify their impact on the knowledge of
the physics at work in these sources.
\end{abstract}

\medskip

\section{Introduction}
The field of Soft Gamma Repeaters (SGRs) underwent a \emph{golden
age} during the operating life of BeppoSAX. This was likely due to
a rare combination of events: availability of suited
instrumentation, effectiveness and ability of the investigators,
and a significant co-operation by the sources. Periodicities were
discovered in the X-ray quiescent counterparts of two SGRs,
providing support to the interpretation of these sources as
\emph{magnetars}, the celestial objects hosting the highest known
magnetic fields. In addition, one source - SGR 1900+14 - emitted
two large flares, one of which similar to the famous 1979 March
5th event. Finally, one new source - SGR 1627-41 - was discovered
by the BATSE experiment (not to mention the 0.5$\pm$0.5 new SGR
that might have emitted just two short bursts and nothing more).

\section{BeppoSAX Statistics}

The BeppoSAX Narrow Field Instruments (NFI) have observed the
three SGR sources that happened to be active during its
operational life time (SGR 1900+14, SGR 1806-20 and SGR 1627-41)
both during active and quiescent periods. In Table 1 we provide
the complete and definitive journal of NFI observations of SGRs,
providing information about the type of observation (Standard or
Target of Opportunity) that also provide a hint on whether the
specific source was active (ToO) or not (Standard) at the time of
the observation. We caution the reader that this works effectively
for the ToO observations, whereas for the standard ones it may
well have happened that the source was, by chance, active at the
time of a well-ahead planned Standard observation.

The total NFI time spent on the SGRs is approximately 735 ks. For
comparison, in Table 2 we list the amount of time spent by the
BeppoSAX NFI on several classes of targets. As can be seen, the
SGRs used only 1.2\% of the total BeppoSAX observing time, and 1\%
of the spacecraft pointings. But in contrast to the small fraction
of observing time spent on these sources, the science return, also
when measured through the number of papers published on refereed
journals making a significant use of the BeppoSAX data, is pretty
large leading so far to an average of $\sim$0.6 papers per
observation, and 0.02 papers per MECS ks.

\vspace{.5cm}
\noindent
\begin{minipage}{13.5cm}
\centerline{\bf Tab. 1 - The BeppoSAX NFI Observations of Soft
Gamma Repeaters}
\vspace{.5cm}

\centering
\begin{tabular}{|c|c|c|c|}
\hline

            &               &               & \\
Target      & Start Time    & MECS Exposure & Observation Type  \\
            &               &               &  \\
\hline
            &                   &               & \\
SGR 1900+14 & 12 May 1997       & 46 ks         & Standard \\
SGR 1627-41 & 6 August 1998     & 45 ks         & Target of
Opportunity \\
SGR 1900+14 & 15 September 1998 & 33 ks & Target of Opportunity \\
SGR 1627-41 & 16 September 1998 & 30 ks & Target of Opportunity \\
SGR 1806-20 & 16 October 1998   & 31 ks & Standard \\
SGR 1806-20 & 21 March 1999     & 57 ks & Standard \\
SGR 1627-41 & 8 August 1999     & 80 ks & Standard \\
SGR 1900+14 & 30 March 2000     & 40 ks & Standard \\
SGR 1900+14 & 25 April 2000     & 40 ks & Standard \\
SGR 1806-20 & 3 September 2000  & 36 ks & Target of Opportunity \\
SGR 1806-20 & 5 September 2000  & 61 ks & Standard \\
SGR 1900+14 & 18 April 2001     & 46 ks & Target of Opportunity \\
SGR 1900+14 & 29 April 2001     & 57 ks & Target of Opportunity \\
SGR 1806-20 & 6 September 2001  & 50 ks & Standard \\
SGR 1900+14 & 27 April 2002     & 83 ks & Standard \\
            &                   &       &  \\
\hline
\end{tabular}

%\smallskip
%$^a$ IAU name; $^b$units of 10$^{-5} \mu$Jy; $^c$ Radio Loud = $P_{5GHz} > 5\times 10^{24}$ watt Hz$^{-1}$ \\
\vspace{0.3cm}
\end{minipage}

\noindent
\begin{minipage}{13.5cm}
\vspace{0.5cm} \centerline{\bf Tab. 2 - General BeppoSAX/NFI
Observation Summary$^{*}$ }
\vspace{.5cm} %TO ALLOW SUFFICIENT SPACE BETWEEN THE TEXT AND THE FIGURES
\centering
\begin{tabular}{|c|c|c|}
\hline
                        &               & \\
Source Class            &   Exposure    &    Number of Pointings \\
                        &   (ks)      & \\

\hline
                        &           &         \\
Gamma-Ray Bursts        &    2335   &      67 \\
Stars                   &    4692   &      70 \\
Compact Galactic Sources &   18088  &     578 \\
(\textit{include SGRs})          &           &         \\
SN Remnants             &     3372  &      74 \\
Normal Galaxies         &     1589  &      30 \\
AGNs                    &    21590  &     441 \\
Clusters of Galaxies    &     5106  &      69 \\
Other                   &     1707  &      46 \\
                        &           &         \\
\hline
                        &                   & \\
TOTAL                   &   $\sim$58.5 Msec & 1375 \\
                        &                   & \\
\hline
\end{tabular}

\smallskip
$^{*}$Courtesy of Paolo Giommi \& Milvia Capalbi - ASDC.
\vspace{0.5cm}
\end{minipage}

The SGR sources were also serendipitously observed by the wide
field BeppoSAX instruments, the Wide Field Cameras (WFC) and the
Gamma Ray Burst Monitor (GRBM). The systematic and complete
analysis of the serendipitous SGR observations with the WFC is
still in progress. For the specific source SGR 1900+14, such an
analysis brought (Jean in 't Zand, private communication) to a
total exposure of 2.3 Msec, accumulated over the 6-year BeppoSAX
lifetime: only 9 short bursts were observed, 3 of which were the
precursors of the 18 April 2001 flare (see below). In addition to
that, the WFC detected the early phase of the 18 April 2001 flare
(see below). For the other sources, especially SGR 1806-20 and SGR
1627-41, similar total exposures are expected after a complete
scan of the full WFC data archive, whereas for SGR 0526-66 a
smaller exposure time may be expected being off the Galactic
plane.

On the GRBM side, the effective area of the instrument, peaked
around 200 keV, and the onboard trigger parameters (optimized to
events longer than 1 s) were not optimal for the detection of the
soft short recurrent bursts from the SGR sources. It detected
approximately 20 short events and the two large flares (27 August
1998 and 18 April 2001) from SGR 1900+14 (see below, and Figure
\ref{wfc}).

\section{Main BeppoSAX Results on SGRs: a biased view}

In this section I will outline those that \emph{I consider} the
major results obtained through BeppoSAX in the field of SGRs. I
will go through the results following their chronology, either in
their appearance in the sky or in the literature, without any
attempt to make neither the list or the report on each result
exhaustive. For this reason, we refer the reader to the original
papers for the individual observations.

\subsection{GRBM Detection of the Giant Flare of 27 August 1998 from
SGR 1900+14}

After few years of burst-quiescence, on May 1998 the source SGR
1900+14 resumed to burst activity with the emission of a few short
bursts. An ASCA observation of the quiescent source carried out,
by chance, few days prior to the reactivation revealed a 5.16 s
periodicity in the steady X-ray emission (Hurley et al. 1999).

\begin{figure}
\centering
\epsfysize=11cm % fix the y-dimension and scales x-dim. to y-dim.
%\hspace{0.1cm} \vspace{0.0cm}
 \epsfbox{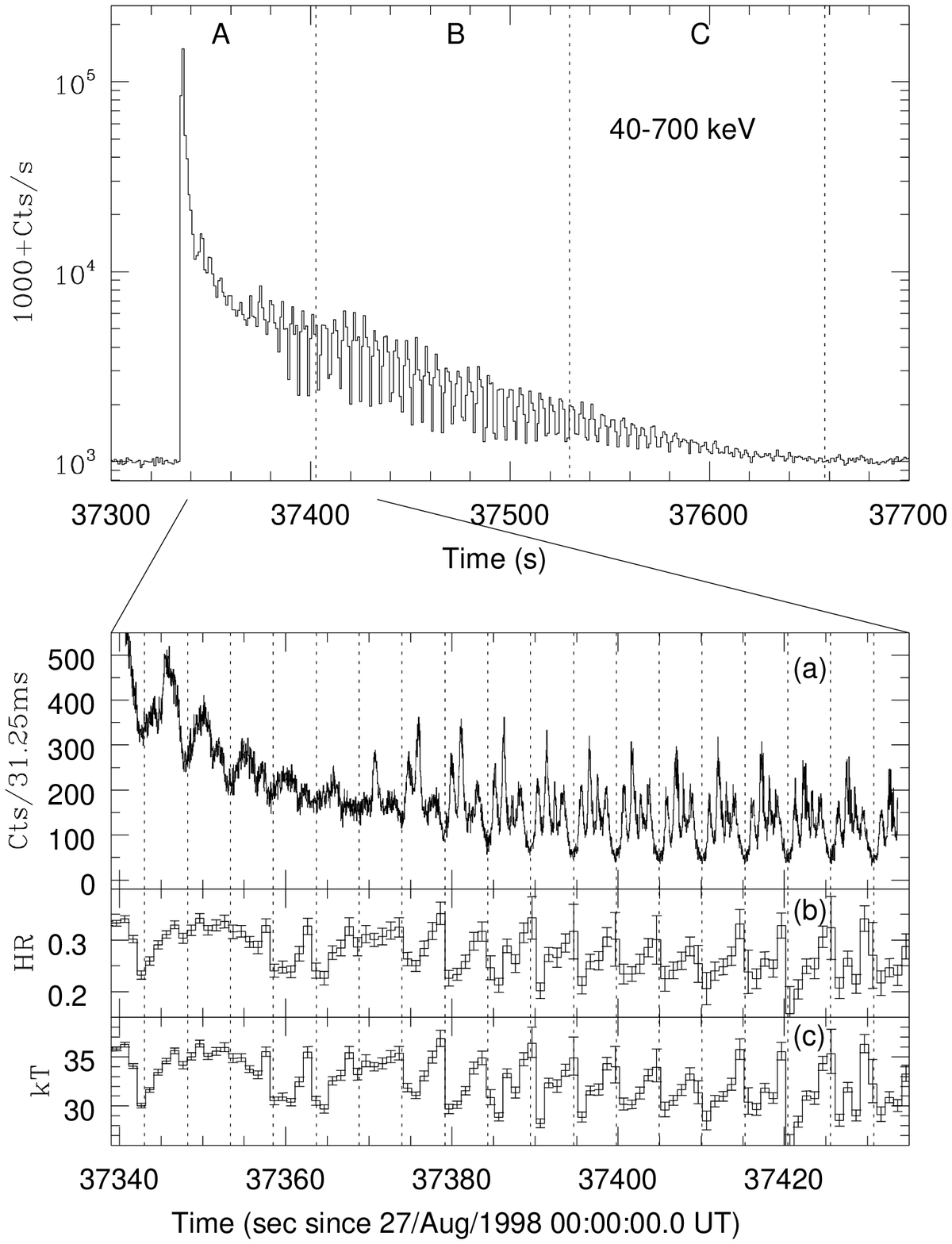}
 \epsfysize=11cm
\centering \epsfbox{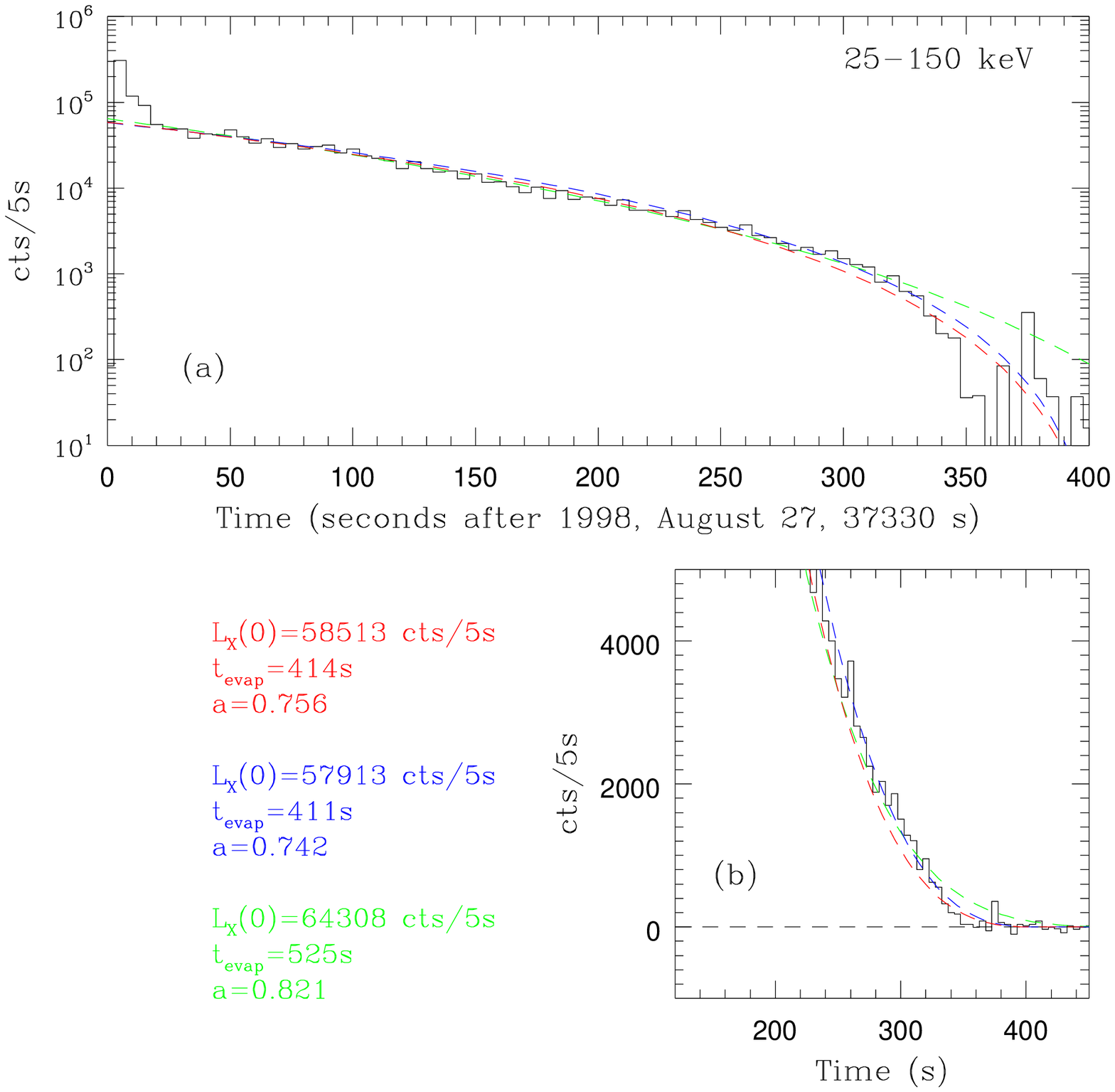}
 \vspace{-0.3cm}
 \caption[h]{\emph{Top}: The giant flare of 27 August
1998 as detected by the BeppoSAX GRBM, together with a 2-channel
spectral evolution expressed in terms of a temperature evolution
(from Feroci et al. 1999). \emph{Bottom}: the light curve of the
same event - here in the 25-150 keV energy range, binned at 5-s to
average over the spin modulation - as described in terms of the
evolution of a trapped fireball with different model parameters
(see text and Feroci et al. 2001a for details).} \label{aug27}
\end{figure}

On 27 August 1998 the source confirmed its new period of activity
with the emission of the second giant flare in the history of the
SGRs (the first one being the 1979 March 5th event from SGR
0526-66). Despite an intrinsic smaller luminosity with respect to
the March 5th event, that of August 27 was the brightest signal so
far ever received at Earth by a cosmic source outside our Solar
System, reaching a peak flux in excess of 3$\times 10^{-2}$ ergs
cm$^{-2}$ s$^{-1}$ above 15 keV (Mazets et al. 1999). The other
exceptional property of this flare was in its soft gamma-ray light
curve: the 5-minutes long exponential-like decay of the flux was
indeed modulated with the same 5.16-s period of the quiescent
pulsar (Figure \ref{aug27}, far top panel). In addition, the data
of the BeppoSAX Gamma Ray Burst Monitor (40-700 keV) first
revealed a very complex (a four-peaked repetitive pattern)
structure of the pulse gradually setting in after $\sim$40 s
(Figure \ref{aug27}, top panel) (Feroci et al. 1999).

A study of this event, carried out using the data from the
BeppoSAX/GRBM and from \emph{Ulysses} gamma ray burst monitor
(25-150 keV) brought to the interpretation and fitting of the
decay curve with the \emph{trapped fireball} model (Feroci et al.
2001a, Thompson \& Duncan 2001). The latter is a somewhat natural
consequence of the interpretation of the giant flares in the
context of the \emph{magnetar} model (e.g., Thompson \& Duncan
1995). In short, this model interprets the giant flare as due to a
magnetically driven instability, involving a large propagating
fracture in the crust of a neutron star hosting a magnetic field
as large as $\sim 10^{14-15}$ G (Thompson \& Duncan 2001). Half or
more of the burst energy is suddenly released in the form of a
relativistic outflow of electron-positron pairs and hard
gamma-rays (the initial hard spike). Part of the fireball is
instead trapped in the magnetic field and evaporates in a finite
time (the \emph{evaporation time} in the trapped fireball
function), by emitting thermal radiation. A fraction of this
thermal radiation (of the order of 20\% in energy) is indeed
Compton-reprocessed to a non-thermal tail by an extended pairs
corona, living around the trapped fireball for the first $\sim$40
s after the initial spike (as indicated by the excess emission
with respect to a pure trapped fireball light curve, Thompson \&
Duncan 2001, Feroci et al. 2001a). As can be seen from the bottom
panel of Figure \ref{aug27}, the trapped fireball model nicely
fits the decaying flux data, especially at the later stages where
it drops much sharper than an exponential (that actually
significantly overestimates the source counts at the end of the
decay curve), at the characteristics evaporation time of the
fireball. The fit is very satisfactory in all the three energy
ranges under study, 25-150 keV, 40-100 keV and 100-700 keV.
Interestingly, the same model describes in a similarly excellent
way the light curve of the March 5th event (Feroci et al. 2001a).

 \subsection{NFI Discovery of a blackbody component in the X-ray spectrum of
the quiescent SGR 1900+14}

Among the properties that the SGRs share with the Anomalous X-ray
Pulsars (e.g., Mereghetti et al. 2002) is the shape of their
energy spectrum. The link has become tighter after the discovery
of a thermal component in the X-ray energy spectrum of SGR
1900+14, obtained by Woods et al. (1999a) using the BeppoSAX/NFI
to observe the source in quiescence. This was the first detection
of a blackbody component ($kT \sim$0.5 keV) in the energy spectrum
of an SGR. Later observations of the same source (e.g., Feroci et
al. 2003, Mereghetti et al., in preparation) show that a spectral
model including a blackbody plus a power law, with interstellar
absorption, is indeed well appropriate to describe the energy
spectrum when the source is in quiescence, whereas the active
source state (that is, during an afterglow - see below) the
blackbody is not requested, meaning that is absent or, more
likely, overwhelmed by the non-thermal spectral component.

Chandra observations of the quiescent counterpart to SGR 0526-66
(Kulkarni et al. 2003) provided indications about the possible
presence of a $\sim$0.5 keV blackbody in the energy spectrum also
of this source, thus confirming and extending the link even
further. The definite proof about the link between SGRs and AXPs,
however, was provided by the recent detection of SGR-like flares
by AXP sources (Gavriil et al. 2002).

\subsection{NFI Discovery and Localization of the X-ray counterpart to SGR
1627-41}

Among the BeppoSAX ''scores" in the field of SGRs, the remarkable
discovery of the X-ray counterpart to the last confirmed SGR,
1627-41, has certainly a primary importance. In fact, after the
first detection of a burst switch-on of this source on June 1998,
its $\sim$ few arcminutes localization was obtained using the All
Sky Monitor onboard RossiXTE, combined with the annulus provided
by the InterPlanetary Network. Starting from these coordinates,
Woods et al. (1999b) used the BeppoSAX/NFI to search for the
quiescent counterpart and found the new source SAX J1635.8-4736,
at a location superposed to the supernova remnant (SNR)
G337.0-0.1. If the SGR and the SNR can be considered physically
associated, then the distance of the SGR would be about 11 kpc,
and the intrinsic source luminosity during the discovery
observation was about 10$^{35}$ erg s$^{-1}$, generally coherent
with the quiescent luminosity of the other sources from this
class.

The location of the new source just on top of the SNR radio
profile (although off-set by the radio core) corroborated the
already claimed association between SGRs and SNRs. However, the
latter association has been more recently argued against, not only
for this source but for the whole class (Gaensler et al. 2001).

\subsection{The Large Flare of 18 April 2001 from SGR 1900+14 }

This section is entirely dedicated to the large flare of 18 April
2001 from SGR 1900+14. For this event BeppoSAX played a major
role, obtaining a series of unprecedented observations. Each of
them is briefly reported in the following sections. For a more
detailed discussion, we refer the reader to the referenced papers.

\subsubsection{GRBM Detection of the Prompt Event }

\begin{figure}
\centering
\epsfysize=12cm % fix the y-dimension and scales x-dim. to y-dim.
\hspace{0.1cm} \vspace{0.0cm}
 \epsfbox{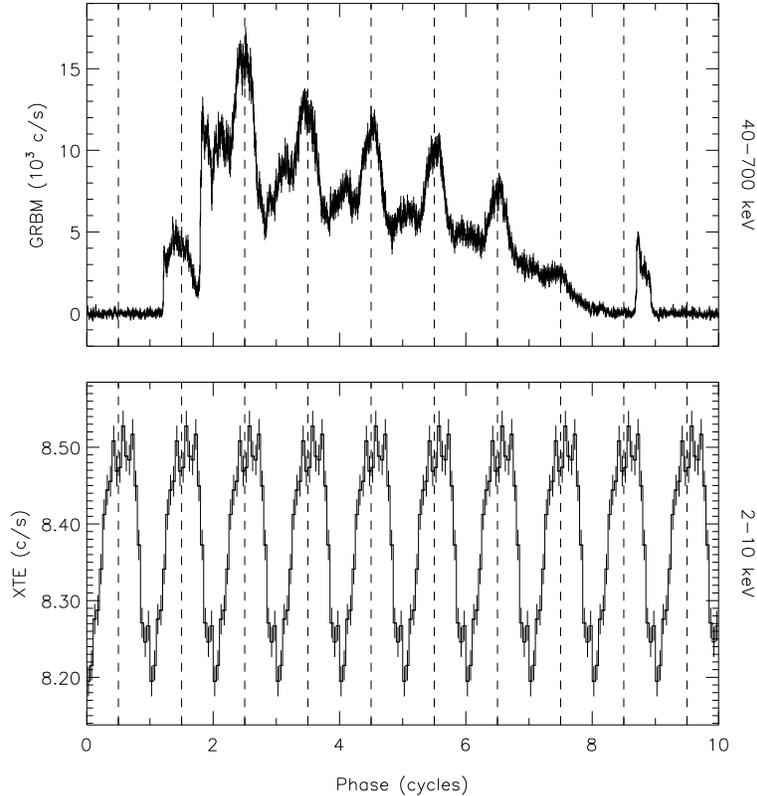}
 \vspace{-0.3cm}
\caption[h]{The 18 April 2001 flare from SGR 1900+14 as detected
by the BeppoSAX GRBM (top panel, Guidorzi et al. 2003) and its
relative timing with the pulsations of the persistent 2-10 keV
emission measured with the RossiXTE PCA few days after the flare
(bottom panel, Woods et al. 2003).} \label{apr18}
\end{figure}

At the time when the large flare went off, 18 April 2001 - 07:55
UT, the Sun was flooding the interplanetary space with a large and
variable flux of protons. For this reason, the two other GRB
detectors that could have provided high quality data of the event,
Ulysses and Konus-Wind, were basically blinded from the solar
particle flux. Therefore, the only instrument that provided high
time resolution data of the event was the Gamma ray Burst Monitor
onboard BeppoSAX (Guidorzi et al., 2001). The top panel of Figure
\ref{apr18} shows the count rate in the 40-700 keV energy range,
with a time resolution of 7.8 ms.

As it appears also from the light curve, this event was
extraordinary under many respects. Similarly to the two giant
flares (5 March 1979 from SGR 0526-66 and 27 August 1998 from SGR
1900+14), the count rate is modulated at the same period as the
quiescent pulsar. In this respect it is worth looking at the
relative timing between the burst light curve and the pulsar, as
derived by observations carried out with the Proportional Counter
Array (PCA) onboard RossiXTE (Woods et al. 2003). This is shown in
the bottom panel of Figure \ref{apr18} that shows how the first 7
pulses of the burst are consistent with being in phase with the
pulsar phase, whereas the last peak is the only one off-set and
seems almost unrelated with the main event. The timing history of
the pulsar across the April 18 flare is reported in Woods et al.
(2003). It shows that, contrary to what happened for the August 27
giant flare (Woods et al. 2001, Palmer 2002), no detectable glitch
and pulse profile change was induced by this event.

Still on the light curve, two other features are worth to be
noticed. First, the duration of the event. In this case it is
approximately 40 s, whereas in the two giant flare was $>$3
minutes and $\sim$5 minutes. This event, therefore, appears to be
intermediate in duration, breaking the bi-modality that held until
its detection between the short bursts (usually shorter than 1 s,
and never above 4-5 s) and the giant flares (minutes), suggesting
a possible continuous spectrum of durations. Also the energetics
of the April 18 event was intermediate between the short bursts
and the giant flares: the 40-700 keV peak luminosity was $\sim1.3
\times 10^{41}$ erg s$^{-1}$ and the total emitted energy about $2
\times 10^{42}$ erg (assuming a distance of 10 kpc).

The other major difference in the light curve with respect to the
giant flares, is the absence of the \textit{short, very hard peak}
at the beginning of the event. In the case of the August 27 event,
based on the radio (Frail et al. 1999) and gamma-ray (Feroci et
al. 2001a, Thompson \& Duncan 2001) data, this peak was
interpreted as a signature of the ejection of relativistic
particles. If that interpretation was correct, then the absence of
that peak in the April 18 implies that such a mini-fireball
emission did not occurred, or was directed away from the observer.

Although different under many respect, the light curve of this
event shares another property with the two giant flares: the
envelope of the decay can be well described with the same trapped
fireball model that describes exquisitely the light curves of the
two giant flares! In this case the fireball parameters were
t$_{evap} \sim$ 37 s and fireball index $\sim$0.4.

\subsubsection{WFC Detection of the Prompt Event }

\begin{figure}
\centering
\epsfysize=9cm % fix the y-dimension and scales x-dim. to y-dim.
\hspace{0.1cm} \vspace{0.0cm}
 \epsfbox{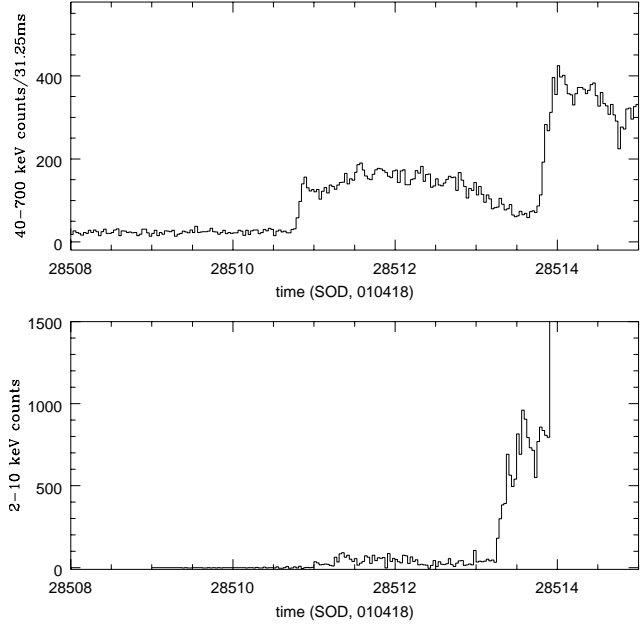}
 \vspace{-0.3cm}
\caption[h]{The 18 April 2001 flare from SGR 1900+14 as detected
by the BeppoSAX GRBM (40-700 keV, top panel) and by the
BeppoSAX/WFC (2-10 keV, bottom panel), before the WFC instrument
was switched off by the self-protective automatic procedure. }
\label{wfc}
\end{figure}

The time of occurrence of the April 18 burst was in a sense very
timely for BeppoSAX. In fact, it occurred at a time when BeppoSAX
was manoeuvring to reach the planned pointing direction for the
NFI. During some manoeuvres, the satellite needed to step through
temporary attitudes.  At the time of the burst the temporary
attitude position was such that SGR 1900+14 was in the field of
view of the WFC! Therefore, this lucky event brought to the first
ever detection of a large SGR flare at soft X-rays. Unfortunately,
the event was so bright that after only 3 s it triggered the
self-protective instrument switch-off, preventing us to collect
X-ray data for the rest of this exceptional event. On the other
hand, the imaging capabilities of the BeppoSAX WFC allowed
immediately to uniquely identify SGR 1900+14 as the emitting
source. The first few seconds of data show an X-ray light curve
significantly different from the simultaneous gamma-ray light
curve, implying a large X-to-gamma spectral evolution (Figure
\ref{wfc}). Instead, using only X-ray data we do not find
significant spectral evolution, nor evident spectral features.

\subsubsection{WFC Discovery of X-ray Precursors }

At the time of the April 18 flare, the source SGR 1900+14 had been
burst-quiet for about two years. Contrary to what happened with
the August 27 giant flare, when the source went out of the
burst-quiescence few months before the flare emitting several
short bursts, in this case no ''standard" short bursts were
detected before the emission of the large flare. Instead, an
inspection of the available pre-burst WFC data revealed 3 short
(100, 125 and 55 ms in duration) and weak bursts, occurred between
2500 and 400 seconds before the main flare (Feroci et al. 2001b).
None of these events was detected by the BeppoSAX/GRBM, or Ulysses
or Konus-Wind, and the search for other pre-flare bursting
activity failed to find any other precursor event in the gamma-ray
data from these experiments. The three events thus represent the
wake up of the source to a burst-active phase, that later
continued with the April 18 flare and with several short bursts in
the following months.

\subsubsection{NFI Discovery of an X-ray Afterglow }

\begin{figure}
\centering
\epsfysize=10cm % fix the y-dimension and scales x-dim. to y-dim.
\hspace{0.1cm} \vspace{0.0cm}
 \epsfbox{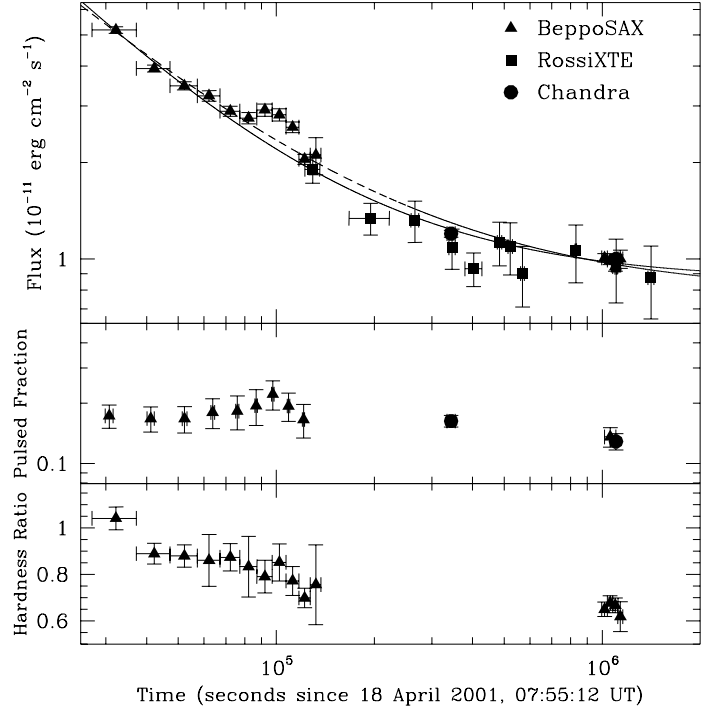}
\vspace{-.3cm} \caption[h]{ \textit{Top Panel}: The X-ray
afterglow detected by the BeppoSAX NFI (full triangles) starting
$\sim$8 hours after the 18 April 2001 flare form SGR 1900+14,
together with the data from RossiXTE (full squares) and Chandra
(full circles). \textit{Middle Panel}: Temporal evolution of the
pulsed fraction, as measured with the BeppoSAX NFI and Chandra.
\textit{Bottom Panel}: Hardness ratio obtained by the (1.6-4) and
(4-10) keV energy bands in the BeppoSAX MECS. (From Feroci et al.
2003.)  } \label{decay}
\end{figure}

Last but not least, BeppoSAX pointed its sensitive narrow field
X-ray telescopes to the source less than 8 hours after the April
18 flare. The pointing lasted for about one day, and was repeated
for the same exposure approximately a week later. The source was
initially detected in 2-10 keV at a flux level $\sim$5 times
larger than the usual quiescent value. As it appears from the
BeppoSAX data points (full triangles in Figure \ref{decay}, top
panel), the 2-10 keV flux decayed by a factor of 5 between the
beginning of the first BeppoSAX observation and the second
pointing, indicating a return to quiescence. Also the bursting
activity accompanied the return to quiescence: several short
bursts were detected during the first observation, whereas a
single burst showed up in in the data of the second pointing. The
data presented here were obtained removing the bursts from the
flux history.

The BeppoSAX observation was also coordinated with a series of
observations with RossiXTE/PCA (squares in the top panel of Figure
\ref{decay}) and Chandra/ACIS-S (circles). The overall set of data
were fit with an analytic law including a constant component -
accounting for an underlying persistent emission - plus a power
law describing the excess (afterglow) emission. As can be seen
from the dashed line in the top panel of Figure \ref{decay}, an
index of 0.89$\pm$0.06 provides a reasonably good description of
the decay trend.

However, on top of the general decay trend, the BeppoSAX light
curve shows a feature at times around $t\sim10^{5}$ s after the
burst. A bump stands out of the power law decay. Any attempts to
attribute this feature to an instrumental effect or to the Sun
activity (that was pretty high during that period) failed, leading
to the conclusion that it was related to the source itself. In the
attempt to characterize the bump, we have carried out a
time-resolved spectral and timing analysis of the BeppoSAX MECS
data, bringing to the results shown in the middle and bottom
panels of Figure \ref{decay} in terms of hardness ratio and pulsed
fraction, respectively.

The spectral analysis reveals a general softening trend of the
2-10 keV radiation. In fact, from the spectral analysis of the two
BeppoSAX individual observations a clear difference is found
(Feroci et al. 2003). The spectrum of the first observation can be
well fit with a simple power law (photon index $\sim$2.6), with
interstellar absorption. Instead, the spectrum of the second
observation requires an additional spectral component, a blackbody
with temperature $\sim$0.6 keV, with the power law becoming harder
(photon index $\sim$1.5). Therefore, the softening observed in the
hardness ratio shows the gradual emergence of the blackbody
component with respect to the power law when the source is
returning to its quiescent status, confirming what already
reported in Sect. 3.2. However, no special spectral signature
seems associated with the bump.

 The timing analysis shows that the pulsed fraction is
generally decreasing, especially from the first to the second
observation (see Woods et al. 2003 for a detailed discussion), and
the bump seems indeed tracked also in the evolution of the pulsed
fraction, although the statistical accuracy does not allow any
clear statement about variability.

\section{Conclusion}

Despite the relative small amount of time spent by BeppoSAX on the
class of Soft Gamma Repeaters sources, a number of unique results
were obtained in this field through its data. These were largely
due to the flexibility of this satellite, both in terms of
instrumentation and managing of the observation program. On one
hand, the large field of view covered with moderate sensitivity by
the WFC and the GRBM allowed to detect unpredictable transient
events, and collect high quality data for them. On the other hand,
the good sensitivity of the imaging narrow field instruments,
combined with their flexible scheduling, allowed to study the
peculiar behavior of these stars soon after their outbursts and to
discover a new counterpart.

I would like to close this review with the content of the last
command uplinked to the satellite on 2002 April 30 at 13:24:52 UT,
after the definitive switch off: \emph{Bravo BeppoSAX}!

\acknowledgements

BeppoSAX has been a 6-year (1996-2002) long successful program of
the Italian Space Agency (ASI) with participation of the Space
Agency of the Netherlands (NIVR). Many people contributed over the
years to its success, from the Instrument Hardware Teams, the
Mission Directors, the Mission Scientist, the teams at the Science
Operation Center, the Operation Control Center, the Science Data
Center, the industrial partners, up to the Science Steering
Committee and the Time Allocation Committees. As a scientist, I
would like to thank them all.

\end{document}